# Exploring Brain Networks Using Noninvasive Electrophysiological Measurements: Methods and Applications

**Richard Leahy & Takfarinas Medani**
University of Southern California

## Foundations of Noninvasive Electrophysiology: EEG vs. MEG

Electroencephalography (EEG) and Magnetoencephalography (MEG) are complementary techniques that non-invasively measure the brain's electromagnetic activity with high temporal resolution [1]. EEG records the electrical potentials generated by neuronal currents through electrodes placed on the scalp. MEG, in contrast, detects the magnetic fields generated by these same neural currents, using highly sensitive magnetometers or gradiometers placed close to the scalp [2]. Both modalities primarily capture synchronous postsynaptic activity in ensembles of cortical pyramidal neurons. Because EEG measures voltage differences, it is highly dependent on the conductive properties of the head (including brain tissue, skull, and scalp), which limits the accuracy with which sources can be localized. MEG measures magnetic fields that are less affected by skull and scalp conductivity and thus can be used to localize neuronal sources more accurately.  A second significant difference is that MEG is particularly sensitive to tangential sources in sulci, whereas EEG detects both tangential and radial source orientations. In practice, EEG tends to have a wider field of spread but can detect deep sources somewhat more effectively. In contrast, MEG has more focal spatial resolution for superficial cortical sources and is less sensitive to radial sources. Both EEG and MEG share the advantage of millisecond-scale timing precision, which enables the tracking of rapid oscillations and cross-regional interactions that underpin cognitive and affective processes. Unlike invasive intracranial recordings (ECoG or sEEG), which directly measure local field potentials within the brain, EEG/MEG are noninvasive and thus more widely used, albeit with lower signal-to-noise and spatial precision than invasive recordings. EEG and MEG can also be recorded simultaneously for complementary information.

**Figure 1.** EEG and MEG measure the electric and magnetic fields generated by synchronous cortical neural activity, which can be approximated as current dipoles arising primarily from dendritic currents in pyramidal neurons. EEG records scalp voltage differences shaped by the conductive properties of the head, whereas MEG detects magnetic fields that are less affected by the skull and are mainly sensitive to tangential cortical sources.

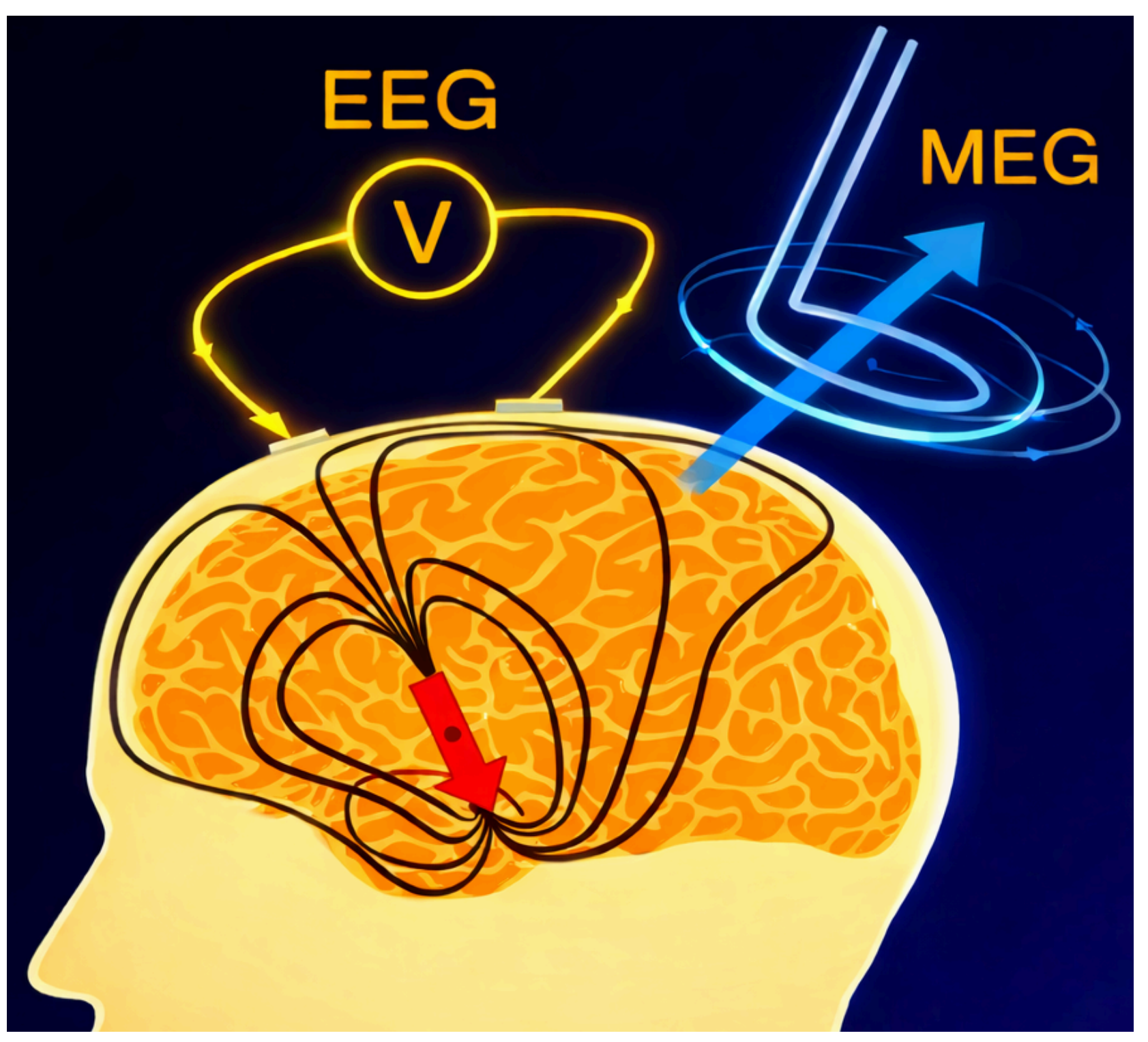

From a network analysis perspective, EEG and MEG data serve as rich sources of input for connectivity estimation [3–6]. However, because they measure field distributions on the scalp or sensor array, interpreting connectivity between sensors is not straightforward. This brings us to a key concept: the need to distinguish sensor-level signals from source-level brain signals. The oscillations captured by one EEG electrode or MEG sensor often originate from multiple underlying neural sources. Likewise, an actual functional connection between two brain regions may not be apparent in sensor recordings, or, conversely, strong correlations between two sensors might merely reflect a single common source influencing both (a volume-conduction effect). Therefore, mapping the brain's networks with EEG/MEG typically requires solving the *inverse problem* to estimate source activity, after first modeling the forward problem of how sources generate the measured fields.

This chapter explores the use of EEG and MEG in reconstructing and analyzing brain networks. We begin with the foundations of noninvasive electrophysiology and the methodological steps required to transform sensor recordings into estimates of the cortical current density, and then compute interpretable network measures. We highlight specific tools and pipelines, many of which are included in the Brainstorm software package (https://neuroimage.usc.edu/brainstorm/), developed at USC in collaboration with the University of Montreal and UTHealth Houston. We review key metrics and approaches in network science as applied to EEG/MEG, including functional connectivity and effective connectivity. We describe advanced concepts, such as time-varying connectivity and cross-frequency coupling, which help link electrophysiological patterns to cognitive functions and emotional states. Throughout, practical guidance and best practices are provided to help researchers avoid common pitfalls (such as volume conduction artifacts) and to embrace data standards (such as BIDS) that enhance the transparency and impact of EEG/MEG network studies.

This chapter is intended for researchers and advanced students in neuroscience, biomedical engineering, and signal processing who seek a principled understanding of EEG/MEG-based network analysis, as well as practical guidance for implementing reproducible pipelines. While the methodological foundations are emphasized, the presentation is application-oriented, aiming to bridge physical modeling, source reconstruction, and network neuroscience within a unified framework.

# From Sensors to Sources: Forward Modeling and Inverse Reconstruction

## The Electromagnetic Forward Problem

The forward problem in EEG/MEG asks: *If a specific neural current distribution is active in the brain, what electric potentials and magnetic fields would we measure at the sensors?* Solving this requires a volume conductor model of the head. Neuronal currents, primarily dendritic currents in pyramidal cells, can be approximated as current dipoles [1,2,7]. The goal of forward modeling is to compute the fields each dipole would generate at scalp electrodes (EEG) or external sensors (MEG). Early solutions modeled the head as concentric spheres with homogeneous conductivities, which allowed analytical formulas. However, such simplistic models ignore the complex geometry and tissue variations of real heads.

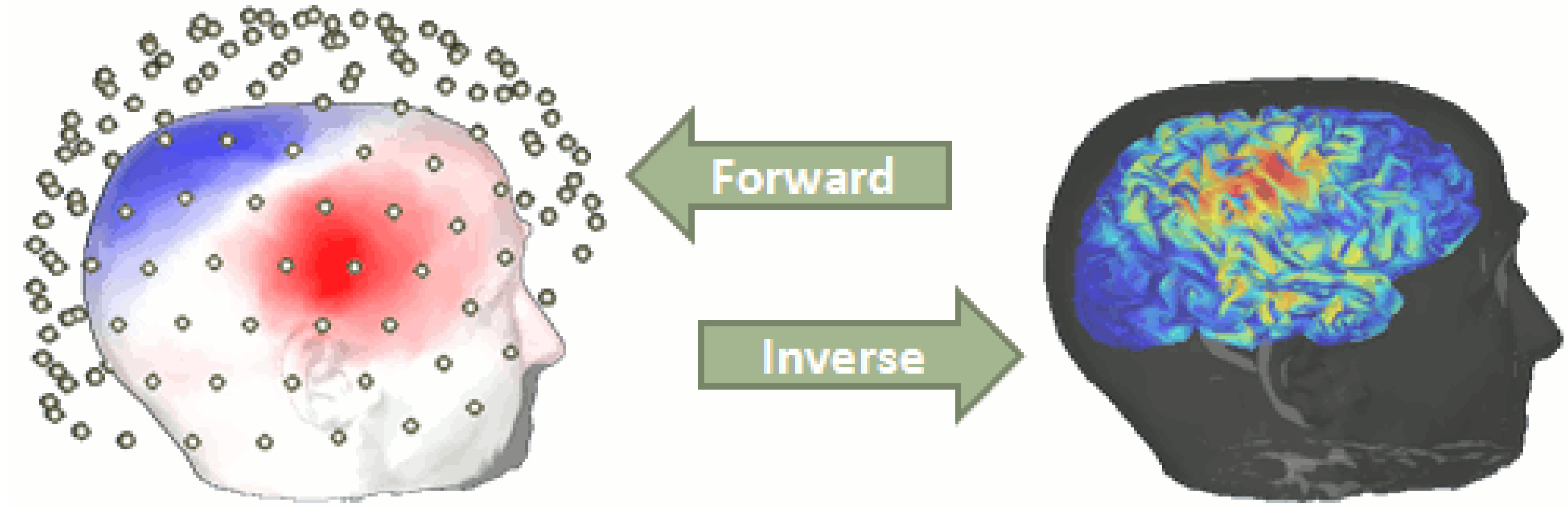

**Figure 2.** The Electromagnetic Forward and Inverse Problems in EEG/MEG. The forward problem maps neural current dipoles in the brain to electric potentials and magnetic fields measured at EEG and MEG sensors using a head volume conductor model. In contrast, the inverse problem estimates the underlying brain sources from sensor measurements.

Modern forward modeling leverages MRI data to build detailed head models that differentiate tissue compartments (e.g., gray matter, white matter, cerebrospinal fluid, skull, scalp, etc.) [8]. This enables more accurate calculation of how currents propagate and attenuate. Two prevalent numerical methods for forward computation are the boundary element method (BEM) and the finite element method (FEM)[7,9–11]. BEM represents tissue boundaries (e.g., inner/outer skull surfaces) with meshes and assumes homogeneous conductivity within each compartment, and computes the forward solution by solving integral equations on the boundaries. BEM is a popular and relatively fast method for creating layered head models, and it is implemented in toolboxes such as OpenMEEG[12]. FEM, on the other hand, subdivides the entire volume into a mesh of elements (tetrahedra or voxels) and can accommodate heterogeneous and anisotropic conductivities within tissues[8,13]. FEM thus allows the use of diffusion MRI to assign anisotropic conductivity to white matter [14,15] (since current flows more easily along than across fiber tracts). FEM is computationally more intensive but increasingly feasible with modern solvers (e.g., the DUNEuro library integrates an efficient FEM solver with the open-source DUNE framework [16,17]). Recent pipelines combine automated MRI segmentation, mesh generation, and FEM solving. For example, a Brainstorm+DUNEuro pipeline can generate a five-compartment head model (white matter, gray matter, CSF, skull, scalp) and compute an FEM forward solution with anisotropic conductivities derived from diffusion-weighted imaging[8,18]. Forward models based on such detailed anatomy have been shown to improve source localization accuracy, particularly for deep sources or when combining EEG and MEG[19,20].

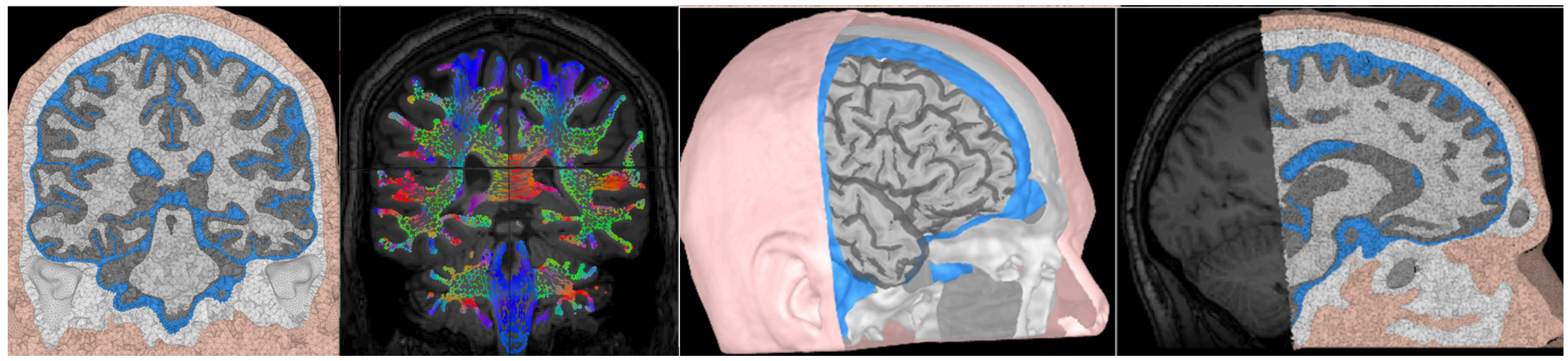

**Figure 3.** Example of subject-specific finite element head modeling for EEG forward simulations. Anatomical MRI segmentation, tissue compartments, and cortical surfaces are combined with diffusion tensor imaging (DTI)–informed conductivity to construct a realistic volume-conductor model for EEG forward computation.

## Template head models vs. subject-specific models is another consideration.

In some studies, especially when individual MRIs are not available, a generic template (such as the MNI/ICBM152 average brain[21] or the USC Brain Atlas[22]) can be used to approximate head geometry[23]. This is expedient, but it introduces errors due to anatomical differences. Subject-specific models using each subject's MRI (or even a structural MRI template matching the subject's age/gender/population) are recommended for best accuracy. Tools like BrainSuite[24] [25], or FreeSurfer (Dale et al. 1999; Fischl 2012), facilitate this by processing MRIs to segment each subject's cortical surfaces and head tissues. For example, BrainSuite implements automated pipelines for skull stripping, tissue classification, and cortical surface extraction, generating outputs (surfaces, volumetric labels) that can then be used for head modeling in EEG/MEG software. BrainSuite's output can be imported directly into Brainstorm and combined with other tools such as Iso2mesh[26] to create BEM surfaces or FEM meshes for the forward model.

## Sensor vs. Source Space, Volume Conduction and Its Impact on Connectivity

Volume conduction refers to the phenomenon in which neuronal current sources generate electrical (volume) currents throughout the head, which are detected almost instantaneously by EEG and MEG sensors [15,27]. In EEG, the low conductivity of the skull causes spatial smearing of the electric potential so that a single cerebral source can produce a spatially broad scalp-potential pattern. In MEG, the physical model differs slightly (for example, MEG is barely distorted by the skull) so that sources can generally be more accurately localized within the brain. The net effect is that signals at two different EEG or MEG sensors may appear highly correlated, even if there is no underlying interaction between two or more distinct brain regions; instead, the sensors might simply be recording the same source's activity. This can create *spurious functional connectivity* at the sensor level[28]. For example, high coherence between two EEG channels could arise from a single strong oscillatory source simultaneously detected by both, rather than from two sources synchronizing. Since volume conduction effects are essentially zero-lag (the same source's signal arrives at multiple sensors with no phase delay due to the 'quasistatic' nature of the underlying electromagnetics), many connectivity measures attempt to discount zero-phase relationships as potentially artifactual[29,30] but instead focus on detecting interactions with non-zero phase delay that cannot be attributed to volume conduction

Several approaches are commonly employed to mitigate the impact of volume conduction or field spread in EEG/MEG connectivity analyses.

- **Sensor Re-referencing and Surface Laplacian:** For scalp EEG, re-referencing can be beneficial. Using a locally constrained reference such as bipolar montage (or applying a surface Laplacian/current source density transform) tends to suppress globally shared components and highlight activity that is spatially focal [28,31]. This is a relatively simple preprocessing step that can reduce the volume-conduction contamination of connectivity estimates, even at the sensor level.
- **Source-Level Analysis:** The most direct approach is to work in source space rather than at the sensors (Schoffelen and Gross,[32]. By applying an inverse solution, we can estimate activity from specific cortical locations rather than relying on mixed scalp signals or magnetometers. This doesn't entirely solve the problem (limited spatial resolution can result in neighboring or correlated sources blurring together, but it does substantially reduce the widespread leakage that occurs at the sensor level.
- **Zero-Lag Metrics:** Another popular strategy is to select a more robust connectivity metric. Measures such as the imaginary part of coherence, the phase lag index (PLI), or its weighted version (wPLI) essentially down-weight or ignore zero-lag correlations [28]. A related technique is to orthogonalize the reconstructed source time series before computing connectivity [33], thereby removing instantaneous coupling that may be artifactual.
- **Experimental contrasts:** Some other studies rely on experimental contrasts[28]. The assumption is that if volume conduction affects both conditions similarly, then comparing connectivity between them (e.g., task versus rest) may primarily emphasize task-related interactions. This is a more indirect approach that depends heavily on the assumption that spatial leakage remains constant across the two states.

Researchers should be aware that no single method completely resolves the issue; volume conduction remains one of the most challenging limitations in noninvasive connectivity analysis. Best practice is to combine complementary strategies, employing accurate source imaging with a well-validated head model, selecting connectivity metrics that mitigate zero-lag leakage, and confirming results through appropriate controls or surrogate analyses.

In the next section, we delve into inverse modeling: the step that takes us from sensor measurements (with volume conduction) to source estimates that enable more meaningful network analysis.

## The Inverse Problem and Source Reconstruction Methods

The EEG/MEG inverse problem is the challenge of reconstructing the distribution of neural activity in the brain from sensor measurements. It is an ill-posed problem: there are infinitely many possible source configurations that could produce a given sensor pattern, so additional constraints or assumptions are needed to obtain a unique solution. Over many decades, various source modeling techniques have been developed [7,34–36]. These can be broadly categorized into *dipole fitting* approaches and *distributed source imaging* approaches:

### Equivalent Current Dipole (ECD) Fitting

In scenarios such as early sensory-evoked responses or interictal spiking in subjects with epilepsy, one might assume a few focal sources. Nonlinear optimization can then determine the dipole parameters (location, orientation, and moment) that best explain the sensor data. Multiple dipole models and localization algorithms can be used to identify two or more dipoles [37,38]. Similarly, beamforming methods use a scanning approach to identify two or more dipolar sources.

**Distributed Source Imaging**

In this case, the brain (often restricted to the cerebral cortex) is divided into a grid or surface of many source locations (referred to as the source space), and the goal is to estimate the current at all these locations that best fit the data. Different techniques impose different constraints.

- **Minimum Norm Estimation (MNE):** This assumes that, among all possible source distributions explaining the data, the one with the smallest overall power (L2 norm) is the most plausible. It leads to a linear inverse solution where source amplitudes are chosen to minimize error while keeping currents as small as possible. MNE tends to produce smooth, spread-out source patterns because it is biased toward many small currents rather than a few large ones[1]. A depth-weighting is often used in the regularizer to reduce bias towards superficial sources.
- **Dynamic Statistical Parametric Mapping (dSPM):** dSPM is a noise-normalized variant of MNE that converts source amplitudes into statistical values by normalizing them with an estimate of the noise variance, typically derived from baseline data. This transformation reduces depth bias and yields maps that are more comparable across brain locations, highlighting statistically significant activations rather than raw current amplitudes. dSPM preserves the linearity and smooth spatial characteristics of MNE while improving interpretability and robustness to noise[39].
- **LORETA and sLORETA:** Low-Resolution Electromagnetic Tomography assume source smoothness; sLORETA (standardized LORETA) further normalizes the solution to reduce localization bias [40].
- **Maximum Entropy on the Mean (MEM):** This approach incorporates Bayesian ideas, searching for the simplest (maximally entropic) distribution of active patches consistent with the data. MEM and related sparse Bayesian methods aim to balance goodness of fit with a preference for spatially sparse or smooth sources, depending on the prior models. These often yield patchy source maps that may more closely reflect true activated regions, though they can be computationally intensive [41].

It is beyond our scope to detail every inverse algorithm; however, the key point for network analysis is that source reconstruction serves as the gateway to estimating brain networks from EEG/MEG[1,7,36]. By obtaining time series for each brain region or voxel, one can compute functional connectivity between those regions, rather than between sensors. This shift to source-space connectivity greatly improves interpretability: a connection can be described as between *brain regions A and B* (e.g., between the left amygdala and the prefrontal cortex) rather than between electrodes F3 and Cz, which is far removed from anatomy and prone to volume-conduction confounds.

**End-to-End EEG/MEG Pipelines: From Sensors to Source Analysis**.

Modern EEG/MEG analysis pipelines increasingly integrate all major processing steps into unified and reproducible workflows. Many tools developed within the research community provide built-in support for source localization, as discussed in the chapter *Software Pipelines and Toolkits for Reproducible EEG Analysis: An Overview*. While the examples given here primarily refer to Brainstorm, other widely used software packages for analyzing EEG/MEG data include FieldTrip [42], EEGLab [43], and MNE-Python [44].

In Brainstorm, a typical EEG analysis workflow includes raw data import; preprocessing steps such as filtering, bad channel detection and interpolation, and artifact removal; coregistration with anatomical MRI when available; and construction of a subject-specific head model using approaches such as boundary element methods (BEM, e.g., OpenMEEG) or finite element methods (FEM, e.g., DUNEuro). This is followed by the computation of the forward model (lead-field matrix), source reconstruction using inverse methods such as minimum-norm estimation or beamforming, extraction of regional time series (e.g., region-of-interest averages or principal component analysis), and subsequent connectivity and network analyses.

These steps can be executed interactively through the graphical user interface or automated using MATLAB scripts, enabling reproducible batch and group-level analyses. The Brainstorm GUI further supports quality control by allowing visualization at each stage of the pipeline, from head model geometry and sensor alignment to source estimates and connectivity results.

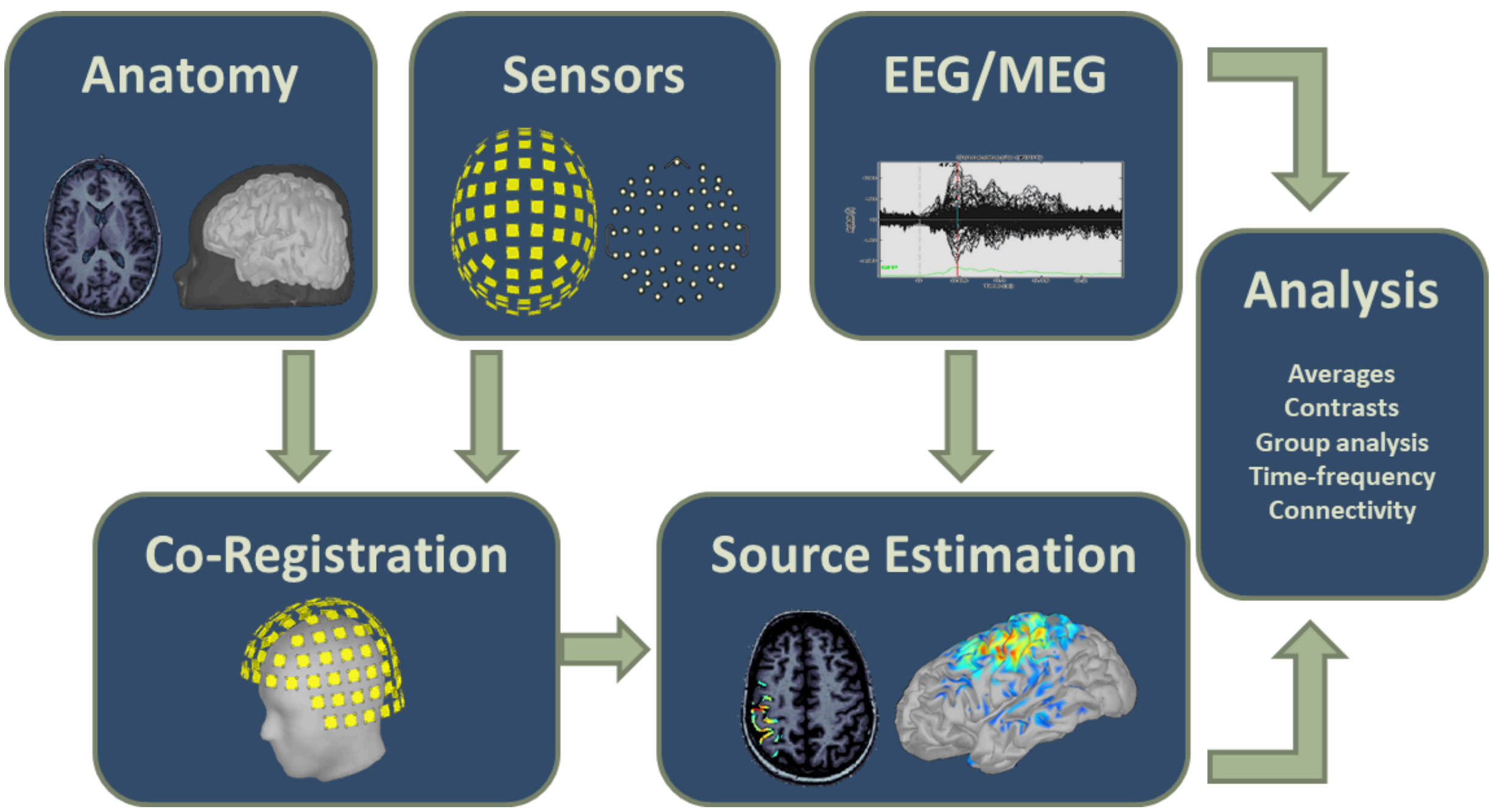

**Figure 4.** Overview of an integrated EEG/MEG analysis pipeline in Brainstorm. The workflow encompasses anatomical modeling, sensor definition and coregistration, preprocessing, forward modeling, source estimation, and downstream analyses, including time–frequency measures and connectivity. The pipeline supports both interactive visualization and automated, reproducible analyses.

## Network Science Approaches for EEG/MEG

Connectivity analyses are commonly performed by computing a bivariate measure between pairs of regional time series of interest. The outcome (a.k.a. a connectome) can be represented as a connectivity graph (left panel below), with each brain region represented as a node with connectivity between them indicated as the edge-strength on the graph between each pair of nodes. The connectome can also be represented by a connectivity array, also known as an adjacency matrix (right panel below). In the case of causal interaction measures, such as Granger Causality, the edges on the graph are directed, indicating the direction of information flow as indicated in Fig. 5.

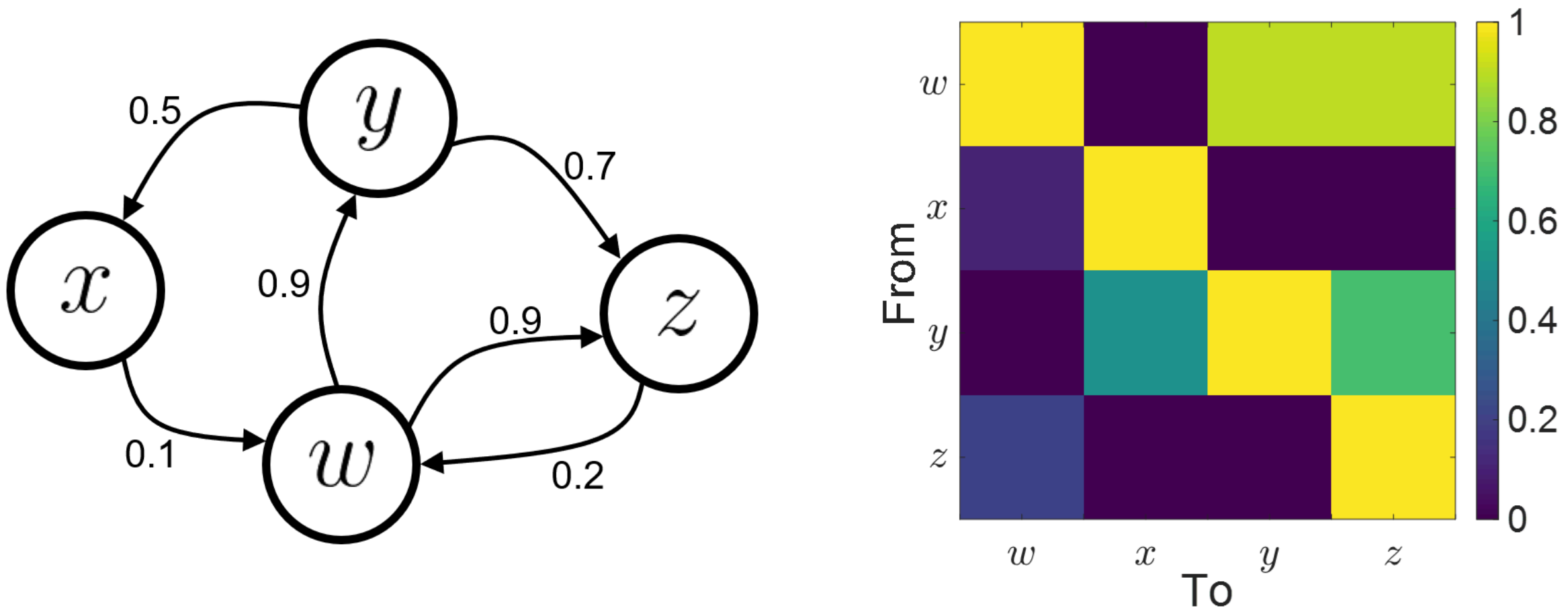

**Figure 5.** Illustration of a directed network and its matrix representation. The left panel shows a weighted directed graph with nodes (w, x, y, z) and edge weights representing connection strengths, while the right panel displays the corresponding adjacency (connectivity) matrix, where rows indicate source nodes and columns indicate target nodes, and color intensity reflects connection strength. Figure adapted from (https://neuroimage.usc.edu/brainstorm/Tutorials/Connectivity)

Having transformed raw EEG/MEG signals into source-level time series, we can now quantify functional interactions and network properties. This section is an overview of the main categories of connectivity analysis as applied to electrophysiological data.

### Functional Connectivity Measures (Undirected)

Functional connectivity (FC) quantifies statistical dependencies between neural signals without inferring the direction of information flow. In EEG/MEG, FC is often estimated in the frequency domain, where oscillatory synchronization reflects a key mechanism for inter-regional communication. Several complementary metrics were used in this study:

- **Coherence and Imaginary Coherence:** Coherence measures frequency-specific coupling by normalizing the cross-spectrum between two signals. High coherence in a given band indicates a stable phase relationship at that frequency. However, conventional coherence is highly sensitive to zero-lag coupling driven by volume conduction. To address this, we also considered the imaginary part of coherence, which retains only the non-instantaneous component of the interaction and therefore provides a more conservative estimate of true functional coupling[28,45].

- **Phase Synchronization Metrics:** Phase-based connectivity is assessed by quantifying the consistency of phase differences across time or trials. The Phase-Locking Value (PLV) detects precise phase synchrony but is also influenced by field spread[46]. In contrast, the Phase Lag Index (PLI) and weighted PLI (wPLI) focus only on interactions with a consistent phase lead–lag relationship, effectively suppressing zero-lag effects[47,48]. These phase-lag measures are now widely recommended for source-space EEG/MEG connectivity and are implemented in tools such as Brainstorm, FieldTrip, and other analysis toolkits[28,49].
- **Amplitude Envelope Correlation:** Connectivity can also arise from co-modulation of oscillatory power rather than phase alignment. Amplitude Envelope Correlation **(**AEC) assesses whether amplitude envelopes fluctuate in tandem over time, capturing slower, large-scale coordination of activity. When applied after orthogonalizing source time series to remove instantaneous shared components, AEC becomes more robust against spurious coupling caused by spatial leakage[28,50].
- **Mutual Information:** For completeness, we note that FC can also be assessed using mutual information, which captures nonlinear statistical dependencies between two signals. However, in EEG/MEG, Mutual Information (MI) is generally less frequency-specific and more susceptible to field spread unless applied carefully in source space.
- **Other Frequency-Domain Measures:** A recent review identified over 40 distinct connectivity metrics in use [51], highlighting that the field has numerous tools and no single “best” measure for all purposes. A wise strategy is to select the connectivity metrics that best test the hypothesis, and to compare multiple metrics to ensure robust conclusions [28,30,52,53].

Any functional connectivity measure will reflect a mix of true neuronal interactions and potential confounds (volume conduction, common inputs, reference effects, SNR issues). Therefore, interpretation is best evaluated by comparing a control with one or more test conditions. Reliability of inference can also be improved by computing connectivity within a network context (e.g., analyzing patterns across the entire connectivity matrix, rather than just single pairs in isolation). Despite these challenges, functional connectivity analysis of EEG has yielded many meaningful insights, although, as with many aspects of EEG/MEG analysis, inference and interpretations should be made conservatively.

**Effective Connectivity (Directed)**

Effective connectivity refers to directed interactions between brain regions[28,54,55], that is, how activity in one neural population influences another over time. Unlike undirected functional connectivity, which only captures statistical association, effective connectivity aims to infer information flow or causal influence within a network. Several methodological frameworks are commonly used in EEG and MEG research:

- **Granger Causality (GC):** GC relies on a predictive modeling principle[56]: if the past of signal *X* improves the prediction of signal *Y* beyond what *Y*'s own past explains, then *X* is said to exert a causal influence on *Y*. In practice, GC is implemented using vector autoregressive (VAR) models in either the time or frequency domain. GC has been applied to EEG source signals to investigate information flow during tasks and resting-state dynamics. However, GC assumes linear, stationary interactions, and both filtering and volume conduction can introduce spurious directionality if not carefully controlled. Consequently, GC is most reliable when applied to source reconstructions rather than sensor-level inference[28,32].

- **Dynamic Causal Modeling (DCM):** DCM employs a biophysically grounded generative model of neural interactions and uses Bayesian model inversion to estimate directed coupling between regions [57,58]. It explicitly represents excitatory/inhibitory influences, as well as synaptic dynamics, enabling the analysis of nonlinear causal relationships. Because computational demands scale steeply with network size, DCM is generally applied to hypothesis-driven models with a limited set of regions (e.g., testing whether prefrontal control over limbic circuits changes during emotion regulation). It offers strong mechanistic interpretability, but is less suited for exploratory whole-brain connectivity mapping.
- **Transfer Entropy (TE):** TE measures the degree to which knowing the past of signal *X* reduces uncertainty about the future of *Y*, after accounting for *Y*'s own history. Unlike linear GC, TE can detect nonlinear and non-Gaussian influences, making it attractive for complex neural dynamics [59]. TE has been used to characterize directed coupling across frequency bands and during state transitions. The main challenges are data requirements and susceptibility to field spread if applied at the sensor level. Source-space estimation and proper selection of the embedding parameter are therefore critical [28].
- **Other Approaches:** A variety of additional techniques exist, including phase-slope index (PSI)[60], conditional mutual information, Bayesian network inference, and causal discovery algorithms (e.g., PC or GES). Many of these methods aim to reduce false causality introduced by common drivers or spatial leakage. Although promising, they typically require careful tuning and remain less established than GC, DCM, or TE in EEG/MEG practice.

Effective connectivity analysis provides a deeper understanding of the network, identifying putative *sources* and *targets* of network interactions. Effective connectivity thus attempts to answer "who drives whom" in the functional networks that EEG/MEG can observe.

# Tools for EEG/MEG Network Analysis: The Brainstorm Ecosystem and Beyond

### EEG/MEG Connectivity Analysis at Sensor and Source Levels within Brainstorm

Brainstorm is a free, open-source application developed initially at the University of Southern California (now a multi-site collaboration) that has become a widely used platform for MEG/EEG data analysis [18]. It provides an intuitive graphical interface for every step of processing, including importing data in various formats, visualizing signals, preprocessing (filtering, artifact correction, and epoching), computing head models and source maps, and performing connectivity and time-frequency analyses. One of Brainstorm's strengths is its emphasis on interactive visual exploration combined with a broad array of computational tools. For example, a researcher can inspect the sensor data, run an ICA to remove artifacts, see the impact immediately, then compute sources and dynamically visualize them on the cortical surface. Brainstorm produces rich graphical outputs for quality checking and result exploration at each stage. This interactive approach lowers barriers for newcomers and facilitates understanding of the data, analysis, and results. At the same time, Brainstorm supports advanced analysis and reproducibility via its scripting capabilities. The entire GUI pipeline includes script analogues, allowing users to generate MATLAB scripts from GUI actions and re-run the same analysis programmatically across multiple subjects or datasets. This is crucial for group studies and for open sharing of pipelines. In a 2019 methods paper, the Brainstorm team presented a comprehensive analysis of a 16-subject MEG/EEG dataset, demonstrating how to interactively analyze a single subject and then automate the pipeline for the group, yielding statistically analyzable results [61]. Brainstorm also uses the increasingly widely adopted BIDS standard for brain imaging data [62].

For connectivity, Brainstorm includes a Connectivity module that offers a variety of metrics for functional and effective connectivity analyses(https://neuroimage.usc.edu/brainstorm/Tutorials/Connectivity). Both bivariate measures (between every pair of signals) and multivariate measures are implemented. Brainstorm can compute frequency-domain measures (coherence, amplitude envelope correlation, PLV, wPLI, etc.), as well as Granger causality and its frequency-domain derivatives (directed transfer function, etc.). Users can compute connectivity either at the sensor level or, preferably, on reconstructed source time series. Brainstorm also supports connectivity-graph analysis. After obtaining a connectivity matrix (for instance, coherence between 68 brain regions), the software can threshold the matrix and display graph metrics, or export the network for further analysis in specialized tools [55]. Visualization of networks in Brainstorm is facilitated by its cortex surface viewer, which allows one to plot connectivity as lines or arcs between brain regions, with strengths indicated by line thickness or color, to obtain an intuitive picture of the network[55].

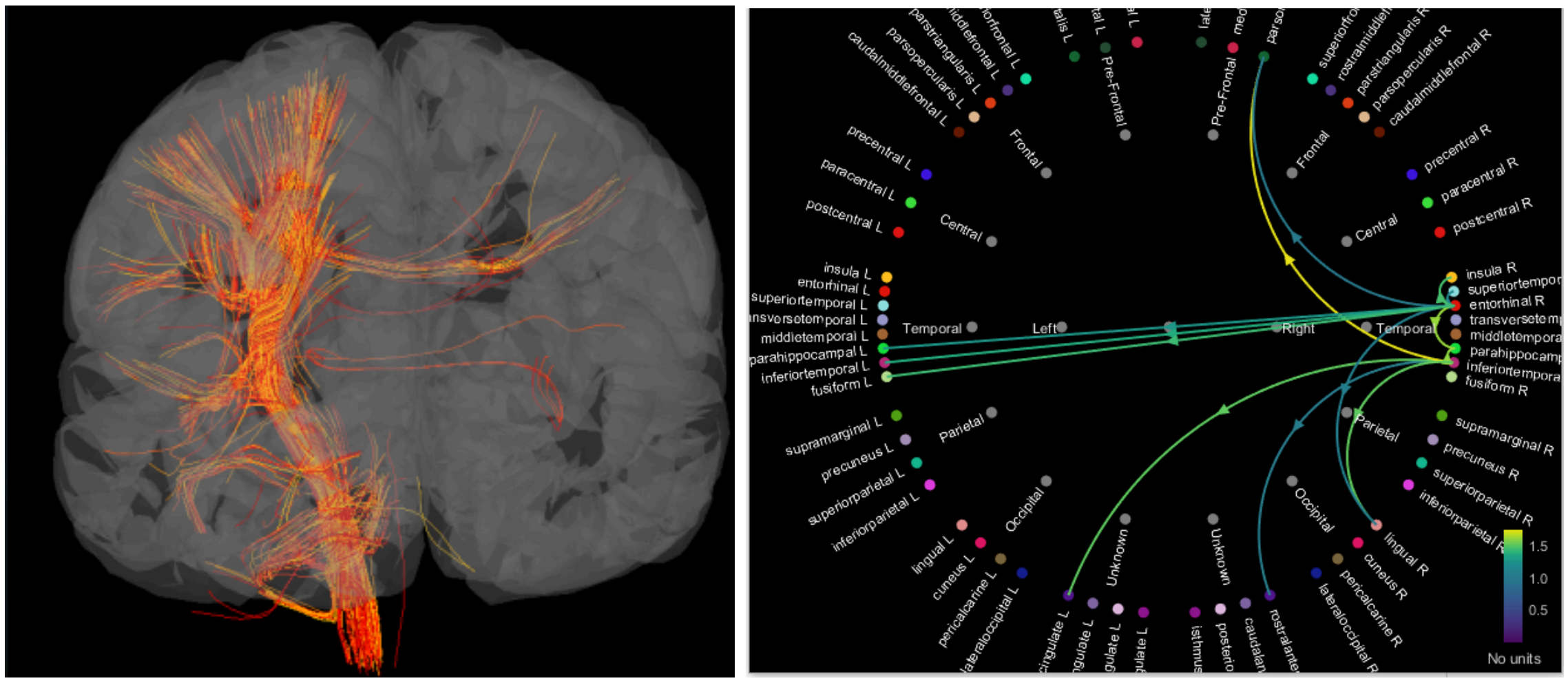

**Figure 6.** Examples of network connectivity visualization in Brainstorm. The left panel displays connectivity mapped onto the cortical surface, where edges between brain regions are rendered as lines or streamlines derived from the white matter fiber tracks, with color and thickness indicating the strength of the connections. The right panel illustrates the same connectivity information using a circular graph representation, facilitating the inspection of inter-regional interactions, hemispheric organization, and network structure.

**BrainSuite for MRI-Based Head Modeling, Structural Connectivity and beyond**

While Brainstorm covers the EEG/MEG processing, it relies on external tools for processing structural MRI data. BrainSuite is an MRI analysis toolkit developed at USC and UCLA[24] that complements Brainstorm by focusing on anatomical data[24]. BrainSuite can take a T1-weighted MRI and automatically perform skull stripping (removing non-brain tissue), tissue segmentation (labeling voxels as gray matter, white matter, etc.), cortical surface extraction (producing a mesh of the gray-white boundary and pial surface), and even cortical atlas labeling. The original BrainSuite paper introduced a pipeline for cortical surface identification that remains at the core of the software. Furthermore, Brainsuite can process DWI and compute the diffusion tensors. Brainstorm can directly read Brainsuite outputs and use them to create head models. Additionally, BrainSuite now provides a BIDS App, enabling it to be run on BIDS-organized datasets in a standardized, reproducible manner [63]. An important feature for EEG users is that BrainSuite can generate high-quality *surface-based* BEM models. The segmented MRI yields the surfaces of scalp, outer skull, inner skull, etc., which can serve as inputs to a BEM solver (such as OpenMEEG, integrated in Brainstorm). Brainstorm also includes dedicated tools for accurate EEG sensor registration on the head surface. Electrode positions can be acquired using a 3D digitizer (e.g., Polhemus or FastTrack systems), ensuring precise mapping of sensors to the individual's anatomy. More recently, Brainstorm has introduced support for photogrammetry-based registration using low-cost 3D camera systems. This approach reconstructs the head and EEG cap shapes directly from photographs and can automatically identify and label electrode locations. These features offer a rapid and user-friendly workflow that enhances spatial accuracy compared to standard template-based sensor layouts.